\documentclass[a4paper]{IEEEtran}

\usepackage{mathtools}
\usepackage{float}
\usepackage[center,singlelinecheck=false]{caption}
\usepackage{cite}
\usepackage{graphicx} 
\graphicspath{ {images/} }
\usepackage{epstopdf}
\usepackage{mathptmx} 
\usepackage{times} 
\usepackage{amsmath} 
\usepackage{amssymb}  

\usepackage{nicematrix}

\usepackage[utf8]{inputenc}
\usepackage{amsthm,amssymb}
\usepackage{multirow}
\usepackage{optidef}

\theoremstyle{definition}
\newtheorem{example}{Example}

\newtheorem{definition}{Definition}

\DeclareMathOperator{\bin}{bin}
\DeclareMathOperator{\supp}{supp}

\DeclareMathOperator{\dist}{d}

\newcommand{\wm}{w_{\min}}
\newcommand{\Awm}{A_{w_{\min}}}

\newcommand{\Aiwm}{A_{i,w_{\min}}(\I)}

\newcommand{\I}{\mathcal{I}}

\renewcommand{\H}{\mathcal{H}}

\newcommand{\C}{\mathcal{C}}
\newcommand{\Ci}{\mathcal{C}_i}
\newcommand{\CI}{\mathcal{C}(\mathcal{I})}
\newcommand{\CiI}{\mathcal{C}_i(\mathcal{I})}

\newcommand{\cC}{\mathcal{C}}
\newcommand{\bG}{\mathbf{G}}

\newcommand{\be}{\mathbf{e}}

\newcommand{\by}{\mathbf{y}}

\newcommand{\bc}{\mathbf{c}}

\newcommand{\bv}{\mathbf{v}}

\newcommand{\bzero}{\mathbf{0}}
\newcommand{\bone}{\mathbf{1}}

\newcommand{\bgi}{\mathbf{g}_i}

\newcommand{\bgh}{\mathbf{g}_h}
\newcommand{\bg}{\mathbf{g}}
\newcommand{\bp}{\mathbf{p}}
\newcommand{\bGN}{\boldsymbol{G}_N}
\newcommand{\bu}{\boldsymbol{u}}

\newcommand{\bhu}{\hat{\boldsymbol{u}}}

\newcommand{\ft}{\mathbb{F}_2}

\usepackage{xcolor}
\def\BibTeX{{\rm B\kern-.05em{\sc i\kern-.025em b}\kern-.08em
    T\kern-.1667em\lower.7ex\hbox{E}\kern-.125emX}}

\usepackage{fancyhdr}

\IEEEoverridecommandlockouts                              

\title{
Selective Reverse PAC Coding for Sphere Decoding
}
\author{
\IEEEauthorblockN{Xinyi Gu, Mohammad Rowshan, {\em Member, IEEE},  
and Jinhong Yuan, {\em Fellow, IEEE}}
 \thanks{The authors are with the School of Electrical Eng. and Telecom., University of New South Wales (UNSW), Sydney, Australia. (e-mail: xinyi.gu@student.unsw.edu.au, and \{m.rowshan, j.yuan\}@unsw.edu.au.}
 \thanks{The work was supported in part by the Australian Research Council (ARC) Discovery Project under Grant DP220103596.}

}

\begin{document}

\maketitle
\pagestyle{empty}
\thispagestyle{fancy}
\cfoot{}

\begin{abstract}

Convolutional precoding in polarization-adjusted convolutional (PAC) codes can reduce the number of minimum weight codewords (a.k.a error coefficient) of polar codes. This can result in improving the error correction performance of (near) maximum likelihood (ML) decoders such as sequential decoders and sphere decoders. However,  PAC codes cannot be decoded by sphere decoding. The reason is twofold: 1) Sphere decoding of polar codes is performed from the last bit - due to the lower rectangular shape of the polar transform. Whereas the shape of PAC codes generator matrix is no longer triangular. 2) One may modify the precoding matrix to get a lower-triangular shape. However, this may reduce the minimum distance of the code due to the formation of unwanted cosets. 
This work proposes a selective convolutional precoding scheme with transposed precoding matrix to reduce the error coefficient while avoiding the reduction in the minimum distance.  The numerical results show the improvement of block error rate by 0.2-0.6 dB, depending on the code rate, in medium and high SNR regimes.
\end{abstract}


\section{INTRODUCTION}
\label{sec:intro}
Polar codes \cite{arikan} form a class of capacity-achieving codes. However, they do not provide a satisfactory error correction performance with a relatively low complexity successive cancellation (SC) decoding in finite block length. To address this drawback, SC list (SCL) decoding was introduced in \cite{tal} that provides a near maximum likelihood (ML) block error rate (BLER) at the cost of high computational complexity. Furthermore, it was recently proposed to concatenate convolutional codes and polar codes resulting in a family of codes known as polarization-adjusted convolutional (PAC) codes \cite{arikan2}. PAC codes can reduce the error coefficient of the underlying polar codes due to the impact of convolutional precoding on the formation of minimum weight codewords \cite{rowshan-precoding}. This reduction is expected to improve the performance of PAC codes under ML decoders such as sequential decoders \cite{rowshan-pac1} and sphere decoder. 

It is known that an ML decoder based on an exhaustive search has a complexity of $O(2^K)$, where $K$ is the number of information bits. Sphere decoding (SD) based on the depth-first approach \cite{pohst} reduces the scope of search effectively. It first finds a close candidate solution in terms of Euclidean distance (ED), not necessarily the closest one, to the received sequence. Then, it searches for a closer candidate to the received sequence by tree pruning, if there exists. Sphere decoding was adapted to polar codes in \cite{kahraman} and then several complexity reduction techniques and approaches such as bounded metrics for SD and list SD (LSD)
 were proposed in \cite{guo, 
hashemi, zhou, husmann} to explore the decoding tree structure in search of the optimal or near-optimal path. 

In this work, we propose a selective convolutional precoding scheme in the reverse direction for sphere decoding of polar codes. This scheme significantly reduces the error coefficient of polar codes while avoiding the challenging problem of the reduction in the minimum distance of the underlying polar code. We also analyze the impact of this precoding scheme on the generation of minimum weight codewords in the cosets with respect to the employed convolutional polynomials. The numerical results show that the proposed techniques can improve the block error rate of polar codes under sphere decoding by 0.2 - 0.6 dB. The power gain tends to be larger at high rates and in high SNR regimes. Needless to mention that this performance improving scheme can be employed along with the complexity reduction techniques available in the literature though it is not in the scope of this work. 
\section{PRELIMINARIES}\label{sec:prelim}
We denote by $\ft$ the finite field with two elements. The cardinality of a set is denoted by $|\cdot|$.  
The \emph{weight} of a vector $\be \in \ft^n$ is $w(\be)\triangleq |\supp(\be)|$. The all-one vector $\bone$ and all-zero vector~ $\bzero$ are defined as vectors with all identical elements of  1 or 0, respectively. The summation in $\ft$ is denoted by $\oplus$. Let $[\ell,u]$ denote the range $\{\ell,\ell+1,\ldots,u\}$ and bold letters denote vectors. The (binary) representation of $i \in [0,2^n-1]$ in $\ft$ is defined as  $\bin(i)=i_{n-1}...i_1i_0$, where $i_0$ is the least significant bit, that is $i = \sum_{a=0}^{n-1}i_a 2^a$. A $K$-dimensional subspace $\cC$ of $\ft^N$ is called a linear $(N,K,d)$ \emph{code} over $\ft$ 
if the minimum distance of $\C$, 
\begin{equation}
     \dist(\C) \triangleq \min_{\bc \in \C, \bc' \in \C, \bc \neq \bc'} \dist(\bc,\bc')=\wm, 
\end{equation}
is equal to $d$. We refer to $N$ and $K$ as the \emph{length} and the \emph{dimension} of the code. The vectors in $\C$ are called \emph{codewords}. Note that the minimum weight of a nonzero codeword $\wm$ in a linear code $\C$ is equal to its minimum distance $\dist(\C)$. 

For a binary input additive white Gaussian noise (BI-AWGN) channel at high signal-to-noise ratio (SNR), according to \cite[Sect. 10.1]{lin_costello}, 
the block error rate (BLER) of linear codes under soft-decision maximum likelihood (ML) decoding can be approximated by 
\begin{equation}\label{eq:union_bound}
    P_e^{ML} \approx \Awm Q(\sqrt{2\dist(\C)\cdot R \cdot E_b/N_0}),
\end{equation} 
where $\Awm$ denotes the number of minimum-weight codewords, a.k.a error coefficient because it plays the role of a coefficient for calculating the BLER bound, $Q(\cdot)$ is the tail probability of the normal distribution $\mathcal{N}(0,1)$, and $R$ is the code rate. As $\Awm$ is directly proportional to the lower bound for the error correction performance of a code, it can be used as a measure to anticipate the direction of change in the block error rate when $\Awm$ changes. 

\subsection{Polar Codes and PAC Codes} 
\label{subsec:RMPolar}
Polar codes of length $N=2^n$ are constructed based on the $n$-th Kronecker power of binary Walsh-Hadamard matrix  
$\mathbf{G}_2 = 
{\footnotesize \begin{bmatrix}
1 & 0 \\
1 & 1
\end{bmatrix} }$, that is, $\bGN=\mathbf{G}_2^{\otimes n}$ which we call it {\em polar transform} throughout this paper. We denote polar transform by rows $\bg_i, i\in[0,N-1]$, and elements $g_{i,j}, i,j\in[0,N-1]$. 
A generator matrix of the polar code is formed by selecting a set of rows of $\bGN$. We use $\I$ to denote the set of indices of these rows and $\CI$ to denote the linear code generated by the set of rows of $\bGN$ indexed by $\I$. Note that $\I \subseteq [0,N-1]=[0,2^n-1]$. The characterization of the information set $\I$ for polar codes  relies on the channel polarization theorem \cite{arikan} and the concept of \textit{bit-channel reliability}. A polar code of length $N=2^n$ is constructed by selecting a set $\I$ of indices $i\in[0,N-1]$ with high reliability \cite{arikan}. The indices in $\I$ are dedicated to information bits, while the rest of the bit-channels with indices in $\mathcal{I}^c \triangleq [0,N-1]\setminus \I$ are used to transmit a known value, '0' by default, which are called \emph{frozen bits}. 

To improve the distance properties of polar codes, it was recently suggested in \cite{arikan2} to obtain the input vector $\mathbf{u}=[u_0,\ldots,u_{N-1}]$
by a convolutional transformation using the binary generator polynomial of degree $m$, with coefficients $\mathbf{p}=[p_0,\ldots,p_m]$ as follows: 
\begin{equation}\label{eq:precoding}
    u_i = \sum_{j=0}^m p_j v_{i-j}.
\end{equation} 
This coding scheme is called {\em polarization-adjusted convolutional (PAC)} coding. 
The convolution operation can be represented in the form of upper triangular matrix shown in \cite{rowshan-precoding} where the rows of the  {\em pre-transformation matrix} $\mathbf{P}$ are formed by shifting the vector $\mathbf{p} = (p_0,p_1,\ldots p_m)$ one element at a row. 
Note that $p_0=p_m=1$ by convention. Then, we can obtain $\mathbf{u}$ by matrix multiplication as $\mathbf{u}=\mathbf{v}\mathbf{P}$.

As a result of this pre-transformation, $u_i$ for $i\in\mathcal{I}^c$ may  no longer be frozen (i.e., $u_i\in\{0,1\}$) unlike in polar codes where $u_i=0$ always holds. 
Then, vector $\mathbf{u}$ is mapped to codeword vector $\mathbf{x}=\mathbf{u}\mathbf{G}_N$. Overall, we have $\mathbf{x}=\mathbf{v}\mathbf{P}\mathbf{G}_N$. 

It was analytically shown in \cite{rowshan-precoding,rowshan-err_coef} that by conventional (forward) convolutional precoding, the number of generated codewords with the minimum of weight in the cosets, denoted by $\Aiwm$ for coset $\CiI$, reduces relative to polar codes (with no precoding).

\subsection{Channel Model and Modulation} 
A binary code $\mathcal{C}$ of length $N$ and dimension $K$ maps a message of $K$ bits into a codeword $\mathbf{c}$ of $N$ bits to be transmitted over a noisy channel. The channel alters the transmitted codeword such that the receiver obtains an $N$-symbol vector~$\mathbf{y}$. An ML decoder supposedly compares $\mathbf{y}$ with all the $2^K$ modulated codewords in the codebook and selects the one closest to $\mathbf{y}$. In other words, the ML decoder  finds a modulated codeword $\textup{x}(\bc)$ such that
\begin{equation}\label{eq:likelihood}
    \hat{\mathbf{c}} = \underset{\mathbf{c}\in\mathcal{C}}{\text{arg max }} p\big(\mathbf{y}|\textup{x}(\mathbf{c})\big).
\end{equation}
For additive white Gaussian noise (AWGN) channel with noise power of $\sigma^2_n=N_0/2$, the conditional probability  $p\big(\mathbf{y}|\textup{x}(\mathbf{c})\big)$ is given by 
\begin{equation}\label{eq:cond_prob}
    p\big(\mathbf{y}|\textup{x}(\mathbf{c})\big)= \frac{1}{(\sqrt{\pi N_0})^N}\text{exp} \left(-\sum_{i=0}^{N-1} \left(y_i-\textup{x}(c_i)\right)^2/N_0 \right).
\end{equation}
Observe that maximizing $p(\mathbf{y}|\textup{x}(\mathbf{c}))$ under binary phase-shift keying (BPSK) modulation that maps $c_i\in\{0,1\}$ to $s_i\in\{1,-1\}$, i.e., $s_i=1-2c_i$, is equivalent to minimizing 
\begin{equation*}
    d^2_E=\sum_{i=0}^{N-1}(y_i-\textup{x}(c_i))^2=\sum_{i=0}^{N-1} \left(y_i- (1-2\bigoplus_{j=i}^{N-1}u_j g_{j,i}) \right)^2,
\end{equation*} 
which is called {\em squared Euclidean distance} (SED). Therefore, for $\mathcal{U}=\{\bu:u_i=0,i\in\I^c \text{ and } u_i\in\{0,1\},i\in\I \}$ where $|\mathcal{U}|=2^K$, we have
\begin{equation}\label{eq:euclid_dist}
    \bhu = \underset{\bu\in\mathcal{U}}{\text{arg min }} \big(\mathbf{y}-(\bone-2\bu\bGN)\big)^2.
\end{equation}

\subsection{Sphere Decoding (SD)} 
The sphere decoding is founded based on a simple idea: We search over integer lattice points $\bone-2\bu\bGN$ for valid $\bu$'s that lie in a certain sphere with radius $r$ around a received noisy vector $\by$ in $N$-dimensional space. In the first step of the algorithm, the sphere decoder performs a depth-first search with a very large initialized $r_0$. Radius $r_0$ may reduce later as the decoder explores the decoding tree looking for alternative candidates. By finding the candidate that allows the sphere to have a minimum radius, SD achieves the ML estimation for the transmitted data $\bu$ with respect to the received vector~$\by$.


We adopt the squared Euclidean distance introduced in \eqref{eq:euclid_dist} to determine the likelihood of the followed path. Hence, for the estimated bits $u_i^{N-1}$, 
the branch metric $m_i(u_i^{N-1})$ is 
\begin{equation} \label{eq:path_metrics}
    m_i(u_i^{N-1}) \triangleq \left(y_i-\big(1-2\bigoplus_{j=i}^{N-1}u_j g_{j,i}\big) \right)^2.
\end{equation}
Based on the low triangle form of $\mathbf{G}_N$, SD starts estimating the bit values from level $l = N-1$. The radius $r_0$ then works as a constraint on the path metrics for levels $l = N-1, ..., 0$ so that for candidates within the sphere, we have 
\begin{equation} \label{eq:path_metric_r0}
    \sum_{i=l}^{N-1} m_i\left(u_i^{N-1}\right) \leq r_0^2.
\end{equation}

The first path candidate is found when it reaches $l = 0$ for the first time. Then, the radius is updated to the path metric of the current path, i.e. $r_0^2 = \sum_{i=0}^{N-1} m_i(u_i^{N-1})$. Only one path is followed at a time throughout the decoding. To search for candidates with a smaller SED, the level is moved backward (i.e. $l = l+1$) to explore an alternative solution for $u_{N-1}^l$. 

The branching condition in \eqref{eq:path_metric_r0} is examined by the updated $r_0$ during the path expansion. The paths that fail to meet the condition are located outside the sphere. Regarded as invalid candidates, they are pruned and will not be revisited. Recall that for $l \in \I^c$, $u_l$ is set to 0. Hence, the path is only split when $l \in \I$.

To reduce the computational complexity, path metrics with a lower bound provide a stricter branching condition. Bound $\Lambda_i =  \underset{\bu\in\mathcal{U}}{\text{min}} \big( m_i(u_i^{N-1})\big)$ is given as the minimum branch metric for two possible cases at a certain estimated bit. 
Consequently, the bounded path metrics for branching evaluation can be computed as
\begin{equation} \label{eq:lower_bound_path_metrics}
\begin{aligned}
    \sum_{i=l}^{N-1} m_i(u_i^{N-1}) + \sum_{i = 0}^{l-1} \Lambda_i 
    \leq r_0^2.
\end{aligned}
\end{equation}

The radius $r_0$ is updated once a new valid path is developed over the decoding tree given that the new path has a metric smaller than the current $r_0$. Until all the feasible paths have been attempted by shrinking the radius of the sphere, the tree search completes. The survived path is the demanded ML estimation for $\bu$ that has the minimum SED.


\section{Reverse and Selective Convolutional Precoding} 
Estimating from the last bit $u_{N-1}$ is the prerequisite of decoding with SD to have the generator matrix $\bG$ as a lower triangular matrix such that we get $x_{N-1} = u_{N-1}$. The convolutional precoding performed by $\mathbf{P}\mathbf{G}_N$ does not give a generator matrix in the form of a lower Toeplitz matrix. As a result, we cannot evaluate $x_{N-1} = p_{0}v_{N-m-1}\oplus \cdots \oplus p_{m}v_{N-1}$ in the first step of sphere decoding. In this section, a different approach for convolutional precoding of polar codes for sphere decoding is proposed. 

To preserve the lower triangular matrix after convolutional precoding, the convolution operation is conducted in a reverse direction rather than as in \eqref{eq:precoding}. That is,
\begin{equation}\label{eq:precoding_rev}
    u_i = \sum_{j=0}^m p_j v_{i+j}.
\end{equation}
Observe the difference between \eqref{eq:precoding_rev} and \eqref{eq:precoding} in the subscript of $v$. This converts the upper Toeplitz matrix to a lower one. Matrix $\mathbf{P_r}$ for the reverse convolution can be represented as the transpose of $\mathbf{P}$, i.e., $\mathbf{P_{r}} = \mathbf{P^T}$.
\begin{equation}
\label{eq:conv_gen_rev}
\mathbf{P_{r}} =
\begin{bNiceMatrix}
p_0 & 0 & & & & & \Cdots & 0 \\
p_1 & p_0 & & & & & & \\
\Vdots & \Ddots & \Ddots & \Ddots & & & &\Vdots \\
p_m & \Cdots & p_1 & p_0 & & & & \\
0 & & & & & & & \\
\Vdots & \Ddots & \Ddots & & \Ddots & \Ddots & \Ddots &\Vdots \\
 & & & p_m & \Cdots & p_1 & p_0 & 0 \\
0 & & \Cdots & 0 & p_m & \Cdots & p_1 & p_0
\end{bNiceMatrix}
\end{equation}

The mapping of input vector $\mathbf{v}$ then becomes $\mathbf{x}=\mathbf{v}\mathbf{P_r}\mathbf{G}_N$, making SD a possible solution for decoding PAC codes. To distinguish these codes from PAC codes, we call them \emph{reverse PAC} (R-PAC) codes in the rest of the paper.


However, the minimum distance $d_{min}$ of the code might be reduced by this reverse precoding scheme, leading to the degradation in the error correction performance. To explain the reason, let us define cosets as follows:
\begin{definition} 
Given a set $\I \subseteq [0,N-1]$ for a polar code, we define the set of codewords $\CiI\subseteq \CI$ for each $i\in \I$ in a coset of the subcode $\C(\I \setminus [0,i])$ of $\C(\I)$ as 
\begin{equation}
    \CiI \triangleq \left\{\bgi\oplus\bigoplus_{h\in \H} \bgh \colon \H \subseteq \I \setminus [0,i]\right\}\subseteq \C(\I),
\end{equation}
where the $i$-th row of the polar transform $\bGN$, i.e., $\bgi$, is the coset leader. 
\end{definition}

We know that the minimum distance of a polar code equals to the minimum weight of the rows of the generator matrix or the non-frozen rows in the polar transform. That is, 
\begin{equation}
    d_{min} = w_{min} = \min(\{w(\bgi):i\in\I\}).
\end{equation}
Hence, we use $d_{min}$ and $w_{min}$ interchangeably throughout this paper.
On the other hand, according to Corollary 3 in \cite{rowshan-err_coef}, we~have
\begin{equation}\label{eq:geq_wi}
    w(\bgi\oplus\bigoplus_{j\in\mathcal{H}}\mathbf{g}_h)\geq w(\bgi),
\end{equation}
where $\mathcal{H}\subseteq [i+1,2^{n}-1]$. Hence, the number of the minimum weight codewords denoted by $\Aiwm$, can be formed in the cosets $\CiI$ where the coset leader has weight $w(\bgi)=w_{min}$.

In the forward convolution used in the conventional PAC codes, since for every $\CiI$, we have $i\in\I$, the minimum distance of PAC codes is preserved. However, the backward of convolution in the reverse precoding by \eqref{eq:precoding_rev} might form some coset $\CiI$ where $i\in\I^c$ and $w(\bgi)<w_{min}$. This would reduce the minimum distance $w_{min}$ of the code as per~\eqref{eq:geq_wi}.
\begin{example}
    Suppose we have the polar code (8,4) with $\I=\{3,5,6,7\}$, $\bp=[1\;0\;1\;1]$ and $\bv=[0\;0\;0\;1\;0\;1\;0\;0]$. The minimum distance of this code is $w_{min}=4=w(\bg_i),i\in\{3,5,6\}$. Now, if the reverse precoding process forms a coset where the coset leader is row $\bgi,i\in\I^c=\{0,1,2,4\}$, then according to \eqref{eq:geq_wi}, codeword(s) with weight $w(\bc)<4$ might be generated, because we have $w(\bg_i)<w_{min}$ for $i\in\I^c$.
    For instance, when precoding the frozen bits with indices $i=2,1,0$, we get $u_2=1\cdot0+0\cdot1+1\cdot0+1\cdot1=1, u_1=1,$ and $u_0=1$, respectively.  
    Observe that $\bc= [0\;0\;1\;0\;1\;1\;0\;0]$ and the weight of the code is $w(\bc)=3$, which is less than $w_{min}$ of the polar code. This occurs as a result of having $\bg_0$ as a coset leader and hence according to \eqref{eq:geq_wi}, the weight can be $w(\bg_0)=1$ or larger (here, it is 3).
\end{example}

To avoid the reduction in $w_{min}$, we propose a selective precoding scheme as follows:
\begin{equation}
\text{$u_i$}=
    \begin{dcases*}
        \sum_{j=0}^m p_j v_{i+j} & if $w(g_i) \ge w_{min}$\\
        v_i & otherwise \\
    \end{dcases*}.
\end{equation}
The input vector $\bv$ in Example 1 is then encoded to the codeword $\bc= [1\;1\;0\;0\;1\;1\;0\;0]$, where $w(\bc)= 4 = w_{min}$. 
With selective precoding, $i\in\I^c$ for $w(\bgi)<\wm$ remain frozen such that the cosets are always led by some $i\in\I^c\cup\I,w(\bgi)\geq\wm$. Hence, for all possible codewords, $w(\bc) \ge w_{min}$ is achieved. 

We use the abbreviation SR-PAC for the selective reverse PAC (R-PAC) in the rest of the paper. We use the notion R-PAC($m+1$) and SR-PAC($m+1$) to denote the constraint length, $m+1$, of the convolutional precoding in R-PAC and SR-PAC coding. 




\section{Analysis: Factors Impacting Error coefficient}\label{sec:analysis}


In this section, we analyze the error coefficient $\Awm$ of some codes and how (selective) reverse convolutional precoding of polar codes (in short, R-PAC or SR-PAC coding) affect $\Awm$. Here, we use a method based on the list sphere decoder (LSD) for enumeration.

To obtain the number of the minimum weight codewords, all zero codeword $\bzero$ is transmitted at a very high SNR (e.g., 20 dB). The LSD is employed to find the $L$ most likely paths for the received sequence. In this setup, we are assuming that the $L$ survived paths contain all or most of the minimum weight non-zero codewords given that we choose a large enough list size $L$. Hence, by counting the total codewords with minimum non-zero weight in the list, we can obtain $\Awm$. Out of all codewords counted as $\Awm$, we need to find $\Aiwm$ for every $i$ where $w(\bgi)=w_{min}$. This can simply be performed by classifying the minimum weight codewords in the list based on the  index of the first non-zero bit.


As illustrated in the last section, reverse precoding might introduce codewords with smaller weight than the minimum distance which weakens the codes. By analyzing codes with different rates and various precoding polynomials, we find that for low-rate codes, because of the sparse distribution of information bits, frozen rows with smaller $w_{min}$ have higher chances to be combined with other high-weight rows and become the coset leaders. This results in the reduction of the minimum distance of the codes. This is where SR-PAC coding needs to be employed instead to avoid a reduction in the minimum distance of the R-PAC codes. 

While for high-rate codes, dense distribution of the information bits usually enables reverse precoding to preserve $w_{min}$. In such conditions, applying reverse convolution selectively does not provide a small $\Awm$ as in R-PAC codes. By limiting the row combinations in SR-PAC coding, the error coefficient can not be reduced as many as in R-PAC coding. Hence, R-PAC codes in this case outperform the SR-PAC codes.

Tables \ref{tb:coset_leader} and \ref{tb:coset_leader_reudce_dmin} list $\Aiwm$ for code (64,14) with selective reverse precoding and reverse precoding, where $\bp_7 = [1\;1\;0\;1\;1\;0\;1]$ is used for R-PAC(7) and SR-PAC(7) whereas $\bp_{10} = [1\;1\;0\;1\;1\;0\;1\;1\;0\;1]$ is used for R-PAC(10) and SR-PAC(10). Note that although rows $i=27,39,43,45$ have $w(\bgi=16)$, they are frozen bits in polar coding, hence $\Aiwm$ is undefined for them in Table~\ref{tb:coset_leader}.  

\begin{table} 
\caption{The number of minimum-weight codewords in coset $\Ci$ for SR-PAC(64,14) code  with $w_{min}=16$.}\label{tb:coset_leader}
\setlength{\tabcolsep}{0.5em} 
\centering
\begin{tabular}{|c|c|c|c|c|l} 
\cline{1-5}
\multicolumn{2}{|c|}{ }                     & Polar                    & SR-PAC(7)             & SR-PAC(10)              \\ 
\cline{1-5}
$i$  & $w(g_i)$         & $A_{i,16}$           & $A_{i,16}$           & $A_{i,16}$           &   \\ 
\cline{1-5}
27    & 16               &                      & 16                   & 4                    &   \\ 
\cline{1-5}
39    & 16                &                      & 0                   & 30                   &   \\ 
\cline{1-5}
43    & 16                &                   & 32                  & 18                   &   \\ 
\cline{1-5}
45    & 16               &                 & 16                   & 4                    &   \\ 
\cline{1-5}
46    & 16               & 32                  & 0                    & 4                    &   \\ 
\cline{1-5}
51    & 16               & 64                    & 36                   & 6                    &   \\ 
\cline{1-5}
53    & 16               & 32                   & 16                  & 2                    &   \\ 
\cline{1-5}
54    & 16                & 16                     & 8                   & 1                    &   \\ 
\cline{1-5}
57    & 16                & 16                   & 8                   & 3                    &   \\ 
\cline{1-5}
58    & 16               & 8                      & 4                   & 0                    &   \\ 
\cline{1-5}
60    & 16                & 4                      & 1                  & 1                    &   \\ 
\cline{1-5}
\multicolumn{2}{|c|}{Total}                & 172                 & 137                  & 73                   &   \\ 
\cline{1-5}
\end{tabular}
\end{table}

\begin{table} 
\caption{The number of minimum-weight codewords in coset $\Ci$ for R-PAC(64,14) code with $w_{min}=12$.
}
\label{tb:coset_leader_reudce_dmin}
\centering
\setlength{\tabcolsep}{0.5em} 
\begin{tabular}{|c|c|c|c|} 
\hline
\multicolumn{2}{|c|}{ } & R-PAC(7)    & R-PAC(10)    \\ 
\hline
$i$ & $w(g_i)$                  & $A_{i,12}$ & $A_{i,12}$  \\ 
\hline
37  & 8                         & 0          & 1           \\ 
\hline
40  & 4                         & 3          & 0           \\ 
\hline
41  & 8                         & 4          & 0           \\ 
\hline
48  & 4                         & 5          & 3           \\ 
\hline
\multicolumn{2}{|c|}{Total}     & 12         & 4           \\
\hline
\multicolumn{1}{l}{} & \multicolumn{1}{l}{} & \multicolumn{1}{l}{} & \multicolumn{1}{l}{}
\end{tabular}
\vspace{-20pt}
\end{table}

According to \eqref{eq:geq_wi}, the minimum distance of the proposed codes is determined by the minimum $w(g_i)$ which plays the role of the coset leaders. Precoding reversely brings the main range of leading cosets ahead of the first information bit  of polar codes (in natural ascending order). That is, a coset may be led by a frozen row with an index smaller than the first information bit. Clearly, this happens due to reverse precoding. In the proposed SR-PAC coding, all the coset leaders have $w(g_i) = 16$ so that they can maintain $w_{min} = 16$ as polar codes. While in R-PAC coding, we would have cosets led by frozen rows with $w(\bgi)=8$ that results in a reduction in the weight of the minimum weight codewords down to $w_{min} = 12$ for both R-PAC(7) and R-PAC(10). 
Note that it is not easy to generate rules for the cases when R-PAC codes reduce the minimum distance because it highly depends on the rate profile and convolutional polynomial. we leave this for future research. 

\section{Numerical Results and Discussions} \label{sec:results}

We consider several sample codes for the numerical evaluation of the proposed scheme. The block error rates (BLER) of the (selective) reverse PAC codes of (64, 50) and (64, 14) are shown in Figs. \ref{fig:bler_(64,50)} and \ref{fig:bler_(64,14)}, whereas Fig.  \ref{fig:bler_(128,110)} shows the BLER of the code (128, 110).  
Error correction performance of R-PAC or SR-PAC codes is compared against SC and SCL decoders. 
Let SCLD($L$) 
denote the SCL decoder 
with list size $L$. The polynomials mentioned in Section \ref{sec:analysis} are used for the precoding of R-PAC and SR-PAC (except for the constraint length $m+1=4$ where $\bp_4 = [1\;1\;0\;1]$).  Table \ref{tb:Admin} gives the minimum distance with the corresponding error coefficient for the underlying codes. Note that concatenating these high-rate short codes with a short CRC can result in significant performance degradation due to a large rate loss. 

\begin{table}[ht] 
\setlength{\tabcolsep}{0.6em} 
\renewcommand{\arraystretch}{1.2} 

\caption{Minimum weight $w_{min}$ and the corresponding error coefficient $\Awm$ of polar codes ($^*$ refers to SR-PAC).}\label{tb:Admin}
\centering
\begin{tabular}{|l|l|l|l|l|l|l|}
\cline{1-7}
\multicolumn{1}{|c|}{\multirow{2}{*}{code}} & \multicolumn{2}{c|}{(64, 50)} & \multicolumn{2}{c|}{(64, 14)} & \multicolumn{2}{c|}{(128, 110)} \\
\cline{2-7}
\multicolumn{1}{|c|}{}                      & $w_{min}$ & $\Awm$                  & $w_{min}$ & $\Awm$                  & $w_{min}$ & $\Awm$                    \\
\cline{1-7}
Polar                                          & 4    & 944                    & 16   & 172                    & 4    & 4099                     \\
\cline{1-7}
SR/R-PAC(4)                                     & 4    & 435                    & 16   & 220$^*$                     & 4    & 1621                     \\
\cline{1-7}
SR/R-PAC(7)                                     & 4    & 98                     & 16   & 137$^*$                     & 4    & 240                     \\
\cline{1-7}
SR/R-PAC(10)                                   & 4    & 70$^*$                      & 16   & 73$^*$                      & 4    & 99$^*$                        \\
\cline{1-7}
\end{tabular}
\end{table}


\begin{figure}[ht]
    \vspace{-10pt}
    \centering
    \includegraphics[width=0.9\columnwidth]{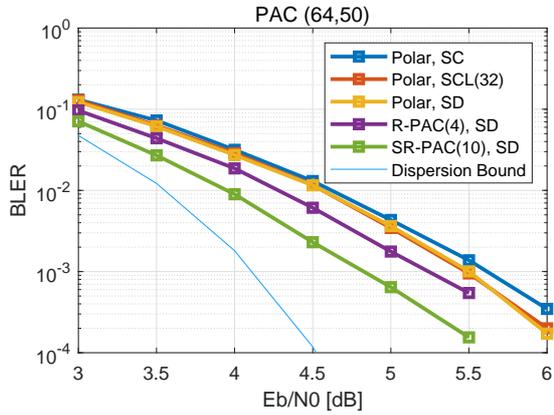}
    \caption{Performance comparison for SR/R-PAC(64,50).}
    \label{fig:bler_(64,50)}
    \vspace{-15pt}
\end{figure}

It can be seen in Fig. \ref{fig:bler_(64,50)} that SD and SCL decoders have the same error performance for high-rate polar code (64,50), while with SD, the proposed R-PAC and SR-PAC outperform polar codes.  From Table \ref{tb:Admin}, we know that $\Awm$ of R-PAC(4) reduces almost by half  compared to polar code. Now relating the error coefficient to the curves, the 0.2 dB power gain in low SNR regimes and a larger gain in high SNR regimes are expected due to a reduction in the error coefficient. As discussed in the last section, longer polynomial results in smaller $\Awm$ for R-PAC and SR-PAC. R-PAC(10) decreases $w_{min}$ of the code, hence we use SR-PAC(10). By employing this code, more than 92.5$\%$ of the minimum weight codewords are eliminated, bringing the curve closer to the dispersion bound \cite{polyanskiy} by achieving up to 0.6 dB power gain. 

\begin{figure}[ht]
    \vspace{-10pt}
    \centering
    \includegraphics[width=0.9\columnwidth]{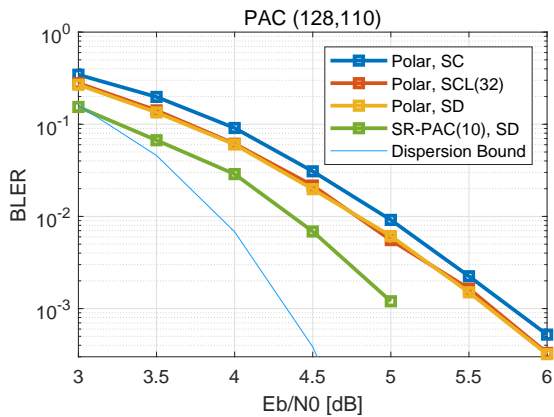}
    \caption{Performance comparison for SR-PAC(128,110).}
    \label{fig:bler_(128,110)}
    \vspace{-10pt}
\end{figure}
Similar conclusions can be drawn for longer codes at high rates, e.g., (128,110). As illustrated in Table \ref{tb:Admin} and Fig. \ref{fig:bler_(128,110)}, the considerable decrease in $\Awm$ for SR-PAC(10) enables the code to outperform its counterpart, polar code.

\begin{figure}[ht]
    \vspace{-10pt}
    \centering
    \includegraphics[width=0.9\columnwidth]{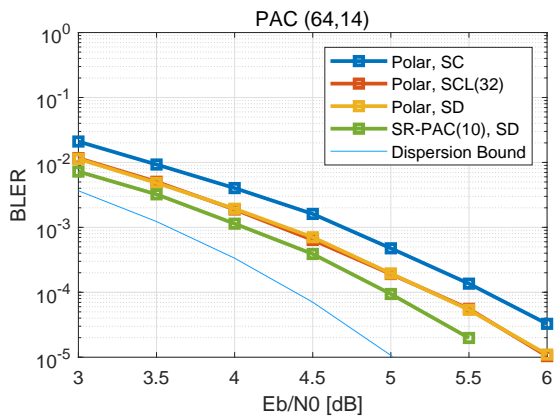}
    \caption{Performance comparison for SR-PAC(64,14).}
    \label{fig:bler_(64,14)}
    \vspace{-15pt}
\end{figure}
For low-rate code (64,14), limited by a relatively small reduction in $\Awm$, the performance gain is not as significant as high-rate codes. As shown in Fig. \ref{fig:bler_(64,14)}, at least 0.2 dB gain can be obtained by the selective reverse precoding scheme. 

\section{CONCLUSION} 
In this paper, 
we propose an approach to concatenate convolutional codes with  polar codes for sphere decoding. The proposed R-PAC and SR-PAC codes have a remarkably smaller error coefficient compared to polar codes. 
This reduction in the error coefficient results in the considerable improvement of the BLER under sphere decoding, in particular for high-rate codes and in high SNR regimes.  
Since this work is focused on improving the code performance by precoding, the techniques to reduce the complexity and design code constructions for each precoding scheme can be considered as the direction of future works. 

\addtolength{\textheight}{-12cm}   







\end{document}